# Transport properties of electrons and holes in a $CuO_2$ layer doped by field effect.

J. Bok and J. Bouvier

Solid State Physics Laboratory - ESPCI. 10, rue Vauquelin. 75231 PARIS Cedex 05, FRANCE.

## Abstract

We propose a model for explaining the recent results obtained by J. Schon et al (1,2) on transport properties of electrons and holes in one plane of $CuO_2$ in a layered cuprate ($CaCuO_2$) where the carriers are created using a field effect transistor (FET). We use the known band structure of a $CuO_2$ plane showing a van Hove singularity (vHs). When the energy of the hole lies near the vHs a variation of the resistivity linear with temperature (T) is calculated and when this energy lies far from the vHs, a quadratic law is obtained at low temperature and becomes linear at higher T. We find that the transition temperature T* is simply related to the distance between the Fermi level and the vHs. The behavior of the Hall coefficient is explained by the existence of hole type and electron type orbits in hole and electron doped $CuO_2$ planes. The fit with the experimental results is excellent.

## Introduction

A recent paper by J. Schon et al (1) reports experimental results on transport properties of electrons and holes in one plane of $CuO_2$ in a layered cuprate ($CaCuO_2$) where the carriers are created employing a field effect device (FET). This device allows to vary the carrier concentration from 0.26 hole per $CuO_2$ to 0.24 electron per $CuO_2$ by applying a voltage on the gate of the FET. The results reported are the variation of the resistivity and Hall cœfficient with temperature, $\rho(T)$ and $R_H(T)$, in this wide range of carriers concentration. This method allows a large variation of carriers concentration without introducing additional disorder linked to the inhomogeneous distribution of chemical dopants and to crystallographic defects. The variation of the hole or electron doping and the exploration of all the phase diagram can be performed also using one single sample. The existence of hole and electron bands can be explained following theoretical band structure calculations made by D. M. Newns et al (3). These calculations use an Hamiltonian containing the p type orbitals of the oxygen atom, the d type orbital of the copper atom, and a large intra-atomic Coulomb repulsion U on the Cu atom. They find that the effect of U is to split the p band in two subbands. The lower Hubbard band can be doped with holes and the upper Hubbard band with electrons. In two dimensions, both bands possess a van Hove singularity. The Fermi level lies close to the vHs for doping levels of the order of 0.20 hole or electron per unit cell.

We present a model that explains all the observed experimental features of $\rho(T)$ and $R_H(T)$ using the band structure for holes in a $CuO_2$ plane as determined by band calculations and confirmed by many experiments such as « angular resolved photoemission spectroscopy » (ARPES) (4-5), and a postulated electron band with analogous properties.

The Fermi surfaces of $CuO_2$ planes, and their variation with the doping as YBCO and the BiSCCO compounds have been studied intensively (4-6), they are well described by the following formula :

$$\varepsilon_k = -2\,t\,(\cos k_x a + \cos k_y a) + 4\,t'\cos k_x a \cos k_y a + (E_F - E_S) \qquad [1]$$

where t is the transfert integral between the first nearest neighbors, t' between the second nearest neighbors, a is the lattice parameter (Cu-Cu distance), $E_S$ is the position of the saddle point (van

Hove singularity (vHs)) and $E_F$ the Fermi level (FL). t and t' have been determined by ARPES in Bi(2212) samples (4-5).
To fit the observed Fermi surface and its variation with hole doping, the following range values have been proposed: t = 0.25 to 0.18 eV and t' = 0.10 to 0.09 eV. The variation of $E_F - E_S$ with doping has been calculated by J. Bouvier and J. Bok (5-6) using the same values for t and t'.

Resistivity

D. M. Newns (7) has shown that the van Hove singularity gives « marginal Fermi liquid » properties when $E_F$ (FL) is very close to $E_S$ (vHs). The lifetime ($1/\tau$) of a quasiparticle is shown to vary as $\varepsilon$ (the energy measured from $E_F$) when the FL lies at the vHs (3-7), and when $E_F$ is far from $E_S$ (more than $k_bT$) the dependence is $1/\tau \propto \varepsilon^2$. This variation is observed experimentally in infrared reflection and in photoemission (3-7).

We use these results to interpret the curves of resistivity reported in the FET experiments (1). For low doping $E_F - E_S$ is large and at low temperature ($k_bT < E_F - E_S$), $1/\tau$ varies as $\varepsilon^2$, and the resistivity $\rho$ as $T^2$. Taking into account a residual resistivity $\rho_o$ due to the impurities, the defects or the interface scattering, we find:

$$\rho = \rho_o + b\,T^2 \qquad [2]$$

for $k_bT > E_F - E_S$, $1/\tau$ varies as $\varepsilon$ and the resistivity $\rho$ as T, we find:

$$\rho = \rho_o + a\,T \qquad [3]$$

with the same $\rho_o$, of course related to a scattering frequency $\nu_o$.
We take $1/\tau_{total} = \nu_o + 1/\tau_{electron\text{-}electron}$.

In our model the temperature T* is the temperature where $\rho(T)$ changes its variation going from $T^2$ to T, so T* is directly related to the distance between $E_F$ (FL) and $E_S$ (vHs). In our theoretical result we find: $E_F - E_S = 1.85\ k_bT^*$. The numerical factor is due to an optimal filling of the DOS at $E_S$ at this temperature.

The calculation of $E_F - E_S$ with the doping is made using the formula [1], and the experimental magnitude of t and t' (t = 0.23 eV and t' = 0.0553 eV for the hole band). We can easily find the doping where $E_F = E_S$, it is the one where the resistivity is perfectly linear in T (from figure 1a of the reference (1)), this hole doping is $p_o = 0.21$. This doping determines the ratio 2t'/t used for our calculations. In Figure 1, we show the fit of experimental curves $\rho(T)$ for various hole doping, and in Figure 2 our calculated T*, reported with the experimental values of reference (1). We can see that the agreement is remarkably good.

In Figure 3, we present several fits of experimental curves $\rho(T)$ for electron doping. We don't show the fits for the underdoped curves, we think that the increase of the resistivity at low temperature, for this doping, is due to a localization and to a mobility edge(8).

Hall effect

The constant energy surfaces for holes in $CuO_2$ planes, given by formula [1] are represented Figure 4. We see that for hole energies (opposite in sign to electron energies) $E < E_S$, the orbits of carriers in a magnetic field are hole like; they give a positive Hall cœfficient $R_H$. For the hole energies $E > E_S$, the orbits are electron like and give a negative contribution to $R_H$.

At very low temperature, the only important orbits are at $E \approx E_F$. The sign of $R_H$ is positive for the hole doping $p < p_o$, and negative for $p > p_o$. The calculated value of $p_o$ is 0.21 hole, corresponding to the experimental value where the resistivity is perfectly linear in T. The opposite signs are true in the electron band.
As the temperature increases, i.e. $k_bT \geq E_F - E_S$, the electron orbits contribute to the Hall effect with a negative sign, then $R_H$ decreases. When the product $\mu_{h,e}B$ is small compared to one, the Hall coefficient is given by:

$$R_H = \frac{1}{e}\frac{p_h\mu_h^2 - p_e\mu_e^2}{\left[p_h\mu_h + p_e\mu_e\right]^2} \qquad [4]$$

where $p_h$, $p_e$ are the density of carriers with hole-like and electron-like orbits respectively, and $\mu_h$, $\mu_e$ their mobility. If $\mu_h = \mu_e$ or B large enough, formula [4] reduces to :

$$R_H = \frac{1}{e}\frac{p_h - p_e}{\left[p_h + p_e\right]^2} = \frac{1}{e}\frac{p_h - p_e}{p^2} \qquad [5]$$

where $p = p_h + p_e$ is the total density of holes. $p_h + p_e$ are calculated using the formula :

$$p_h = \int_{\infty}^{E_S} \rho(E)\,E\,(-\frac{\partial f_{FD}(E)}{\partial E})\,dE \quad , \qquad p_e = \int_{E_S}^{\infty} \rho(E)\,E\,(-\frac{\partial f_{FD}(E)}{\partial E})\,dE \qquad [6]$$

Similar formulae are used for the electron doped case. The results of our calculation are shown in Figure 5, 6 and 7. Figure 5 is for holes. In Figure 6 we plot $R_H / R_H^*$ versus $T/T^*$ and find a universal curve, as found experimentally (1). This is natural because our only variable is $E_F - E_S$ when we change the doping. In Figure 7, we present the calculated Hall coefficients for electron doping. At low temperature and low doping we can account for the maximum observed, using a mobility edge in our calculation. This mobility edge here is the bottom of the upper band. The comparison with experiments gives the general variation of $R_H$ and its universal character, but doesn't fit exactly the details.

## Conclusion

In conclusion, we have shown that all the experimental results obtained by J. Schon et al on the resistivity and Hall effect of electrons and holes in one plane of $CuO_2$ in a layered cuprate can be explained using the band structure of this compound in the framework of the van Hove scenario and without evoking a "pseudogap". This single doped monolayer is a perfect object to check the van Hove scenario in the HTSC cuprates. J. Bouvier et al (9) have already proposed the same explanation for the maxima observed in the variation with temperature of several measured quantities: Pauli susceptibility, the specific heat, the Knight shift, the thermoelectric power etc…(10). On the other hand, some experiments (11) show a pseudogap, i.e. a loss of states near the Fermi level. This pseudogap may be related to disorder introduced by doping (12). In some highly disordered samples two characteristic temperatures T° and T* are observed corresponding to these two different effects.

This paper is devoted to the calculation of the transport properties in the normal state (T>Tc). J. H. Schon et al (1) have also measured the variation of Tc with doping. This was already calculated by J. Bouvier and J. Bok (13) in the framework of the van Hove scenario. We think that the asymmetry of Tc between electrons and holes is due to different screening strengths and coupling (9).

Acknowledgements : We are grateful to Catherine Deville Cavellin for illuminating discussion.

References

(1) “Mapping the Phase Diagram of electrons and Holes in a Layered Cuprate Using Field-Effect Doping.” Preprint (private communication), by J. H Schon, M. dorget, F. C. Beuran, X. Z. Xu, M. Laguës, C. Deville-Cavellin, C. M. Varma and B. Batlogg.
(2) "Superconductivity in CaCuO2 as a result of field-effect doping" J. H. Schon, M. Dorget, F. C. Beuran, X. Z. Xu, E. Arushanov, C. Deville Cavellin, and M. Laguës, *Nature* **414**, 434 (2001).
(3) D. M. Newns, P. C. Pattnaik and C. C. Tsuei, *Phys. Rev. B* **43,** 3075 (1991).
(4) M. R. Norman, H. Ding, M. Randeria, J. C. Campuzano, T. Yokoya, T. Takeuchi, T. Takahashi, T. Mochiku, K. Kadowaki, P. Guptasarma and D. G. Hinks. *Nature* **392**, 157 (1998).
(5) See for a review : “Superconductivity in cuprates, the van Hove scenario : a review.” J. Bouvier, J. Bok. “The Gap Symmetry and Fluctuations in HTSC.” Edited by Bok et al., Plenum Press, New York, 37 (1998).
(6) J. Bouvier and J. Bok, *Physica C* **288**, 217 (1997).
(7) D. M. Newns, *Comment Cond. Mat. Phys*. **15**, 273 (1992).
(8) M. Osofsky et al, under writing.
(9) J. Bouvier and J. Bok, *J. of Supercond*. **10**, 673 (1997).
(10) J. R. Cooper and J. W. Loram, *J. Phys. I France* **6**, 2237 (1996).
(11) See for a review T. Timusk and B. Statt, *Rep. Prog. Phys.* **62**, 61 (1999).
(12) J. Bouvier and J. Bok, *Physica C* **364-365**, 471 (2001).
(13) J. Bouvier and J. Bok, *Physica C* **217-225**, 217 (1997).

Figure captions :

Figure 1: ρ(T) for hole doping
full lines : experimental curves (1)
dashed lines : fit for some curves

Figure 2: full line : our calculated T* compared with the experimental values (dots), for hole doping.

Figure 3: ρ(T) for electron doping
full lines : experimental curves (1)
dashed lines : fit for some curves

Figure 4: energy surfaces in $CuO_2$ planes, given by formula (1), and representation of the hole or electron like orbits.

Figure 5: Calculated Hall coefficient $R_H$ for hole doping using formula [5].
With t = 0.23 eV and t' = 0.0553 eV

Figure 6: Universal curve for the ratio of $R_H/R_H^*$ (here we cannot have true value for $R_{H\infty}$).

Figure 7: Calculated Hall coefficient $R_H$ for electron doping using formula [5].
With t = 0.18 eV, t' = 0.06057 eV, 2t'/t = 0.673

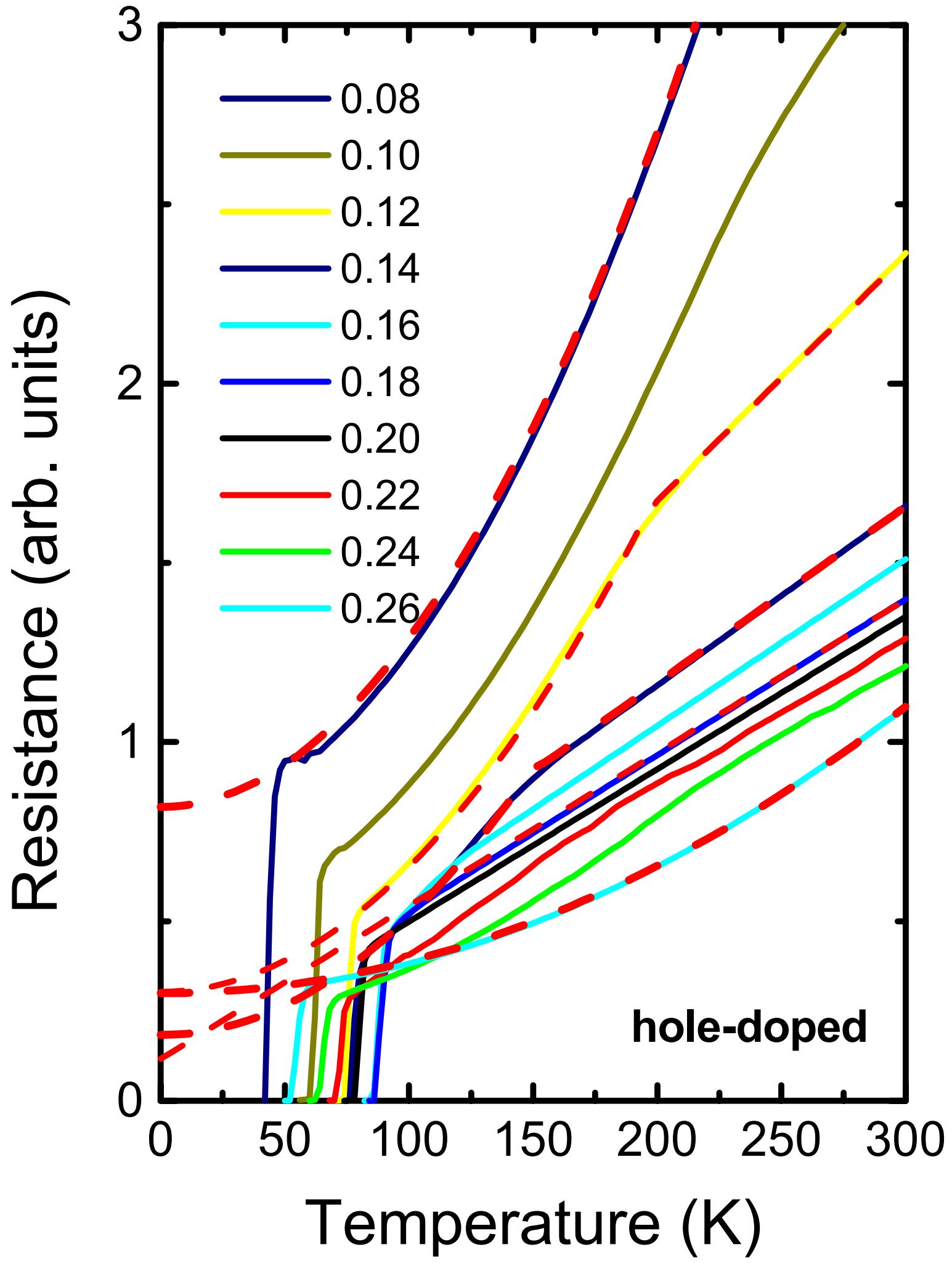

Figure 1: ρ(T) for hole doping
full lines : experimental curves (1)
dashed lines : fit for some curves

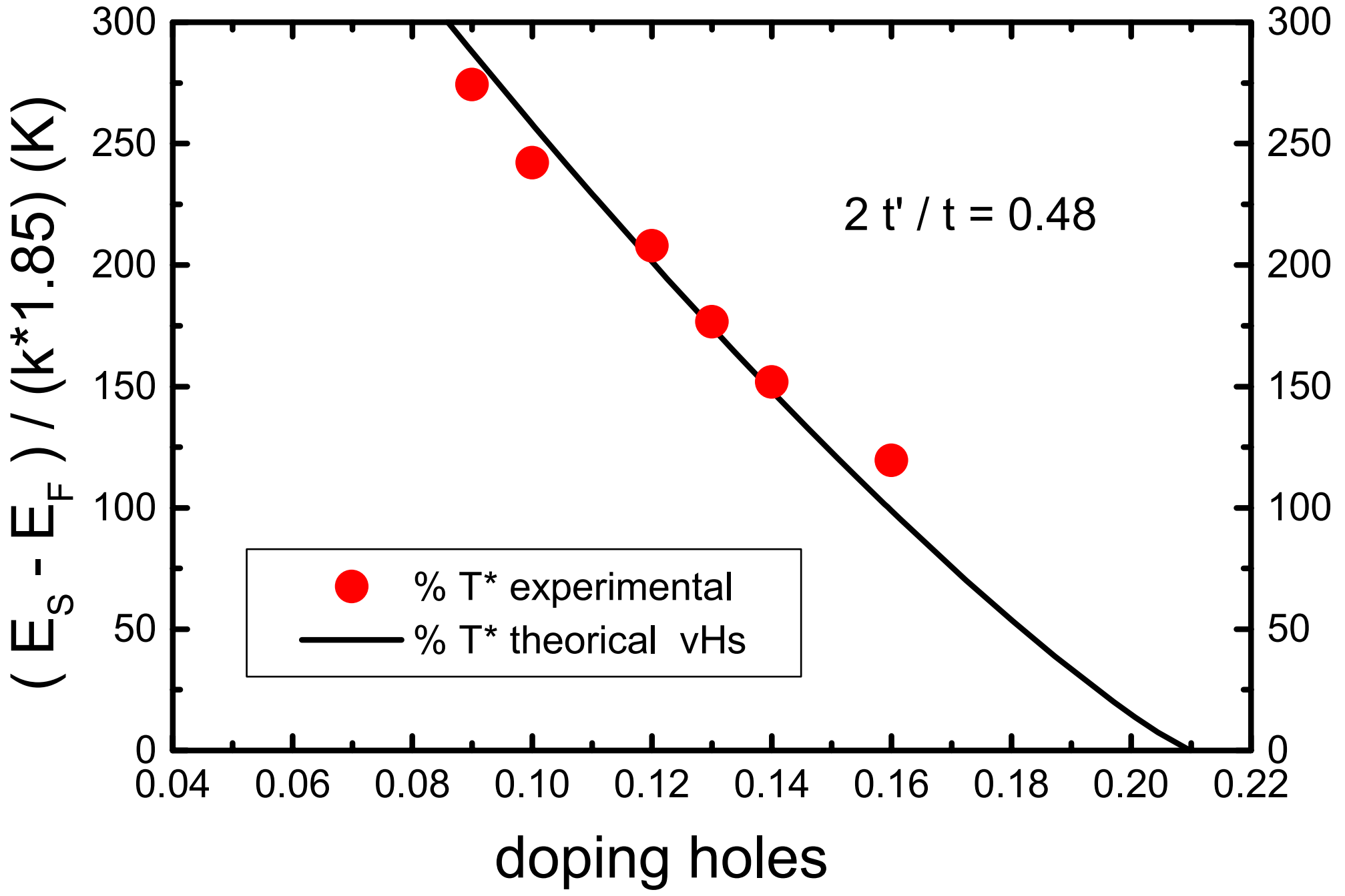

Figure 2: full line : our calculated T* compared with the experimental values (dots), for hole doping.

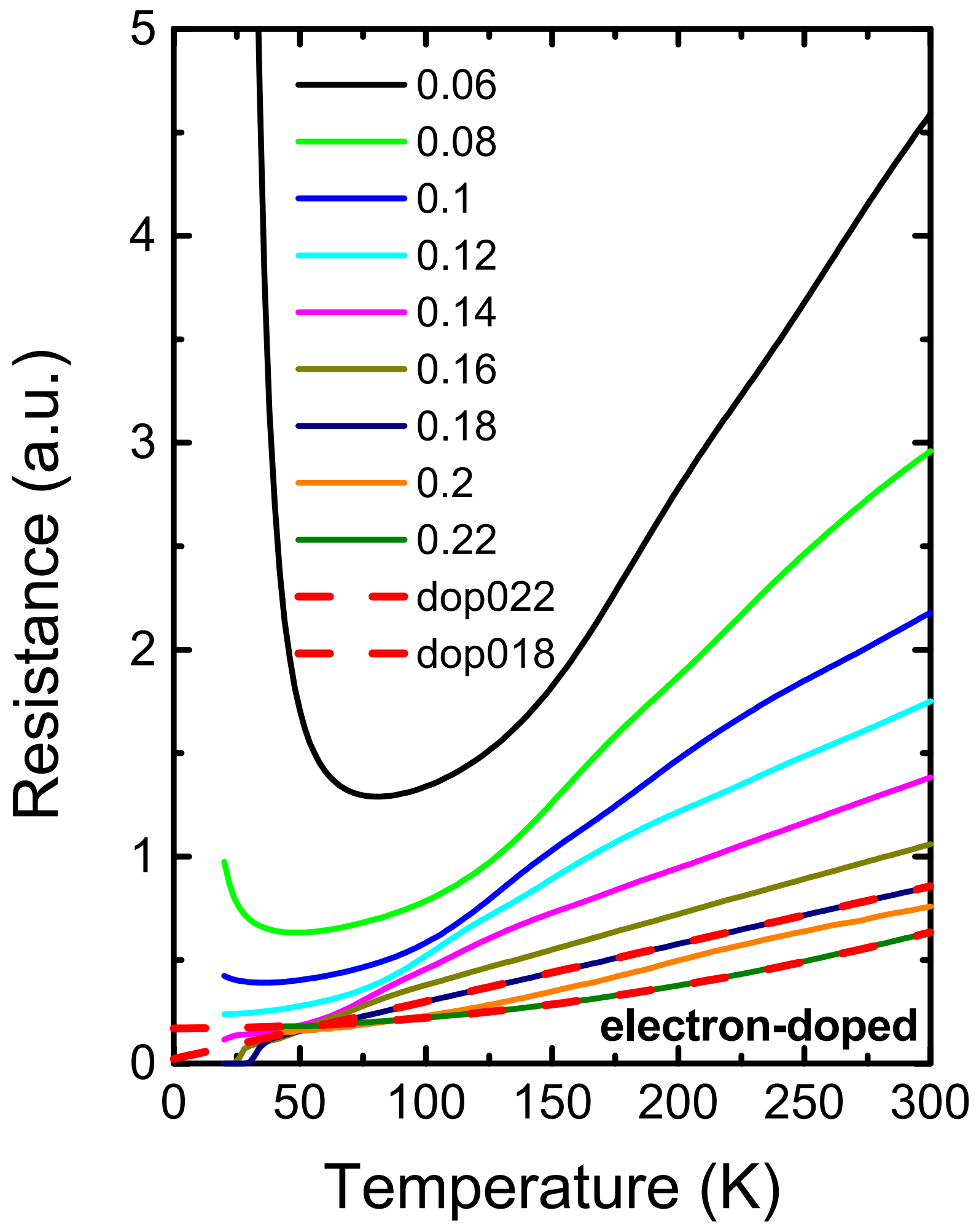

Figure 3: ρ(T) for electron doping
full lines : experimental curves (1)
dashed lines : fit for some curves

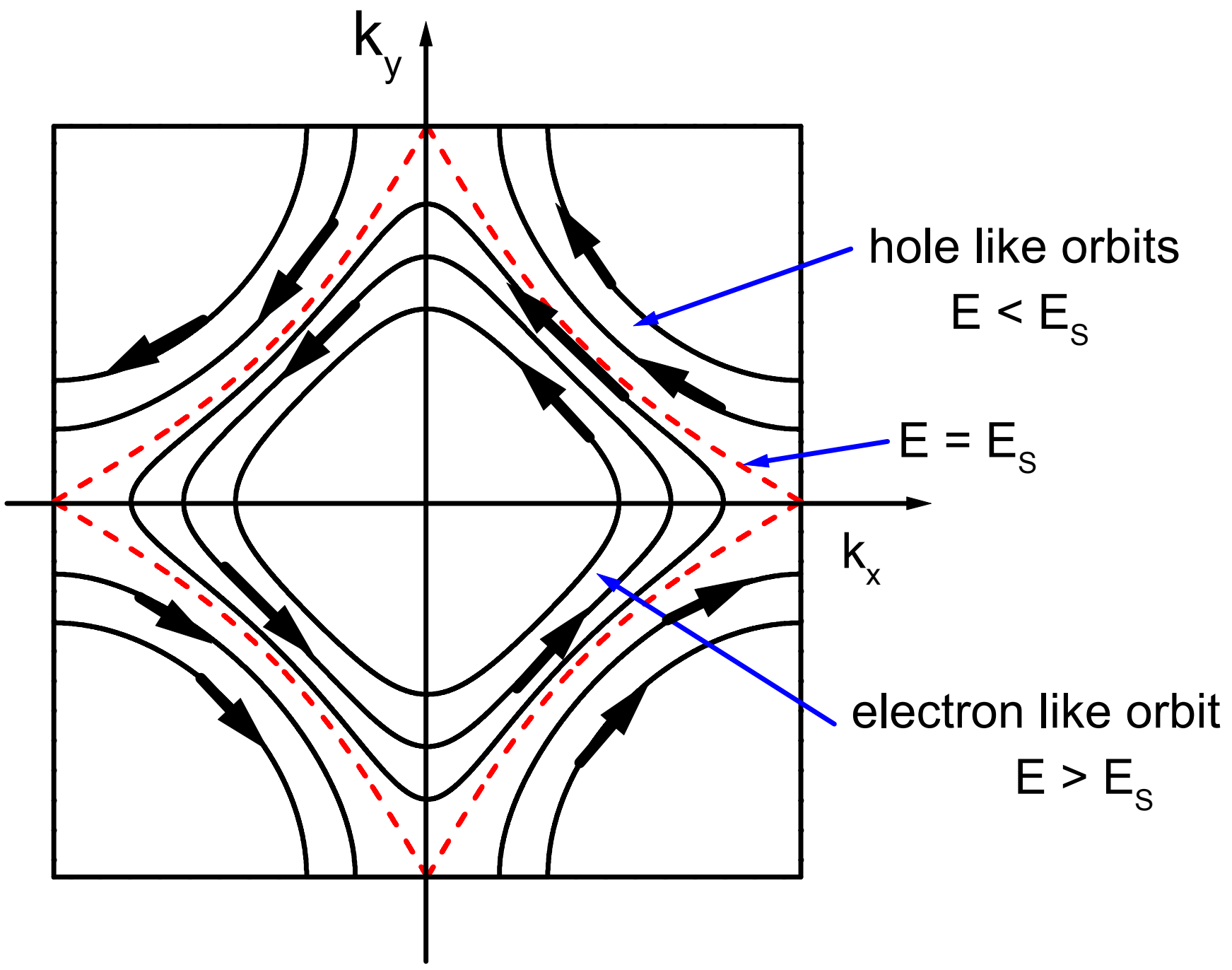

Figure 4: energy surfaces in $CuO_2$ planes, given by formula (1), and representation of the hole or electron like orbits.

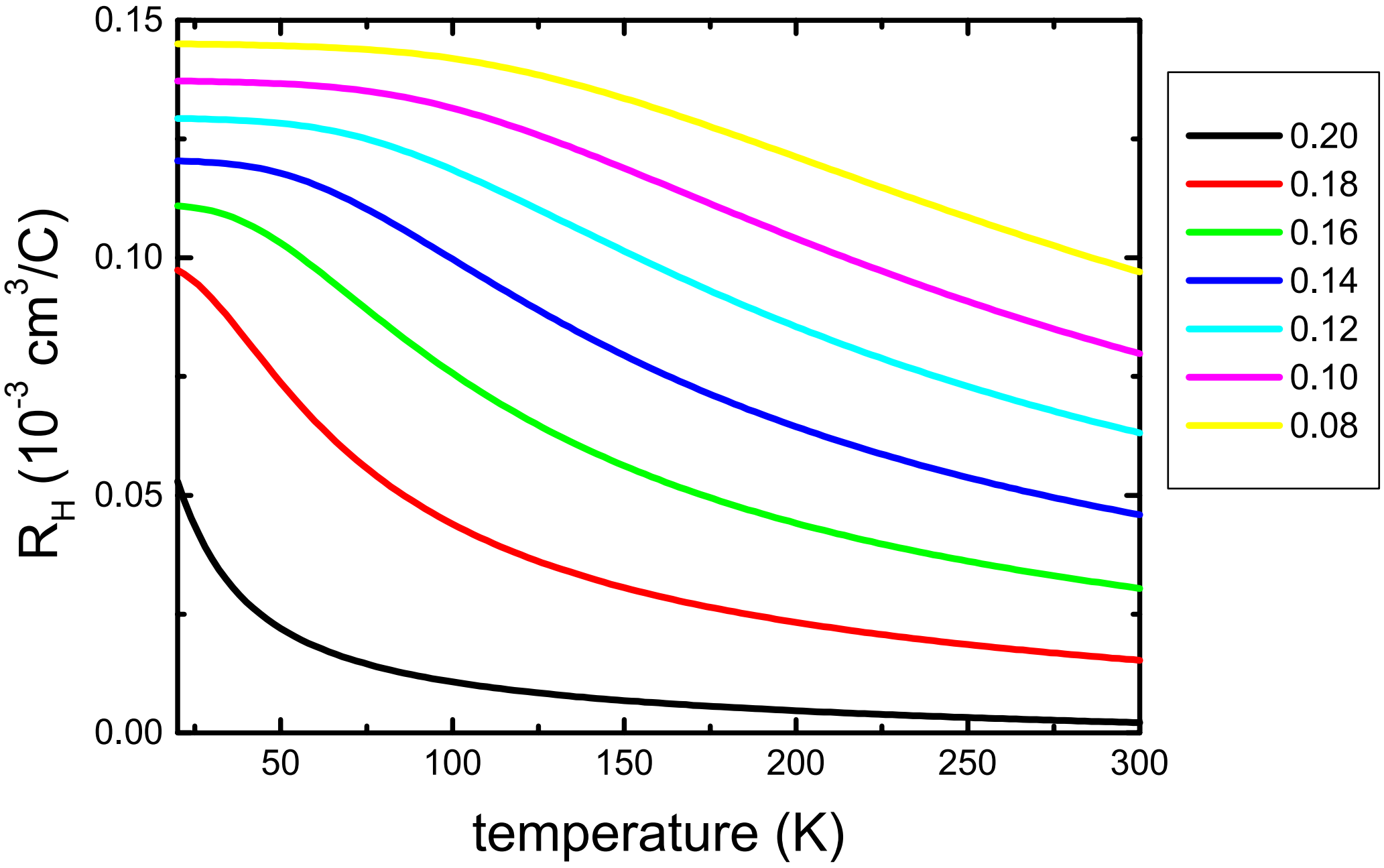

Figure 5: Calculated Hall coefficient $R_H$ for hole doping using formula [5].
With t = 0.23 eV and t' = 0.0553 eV

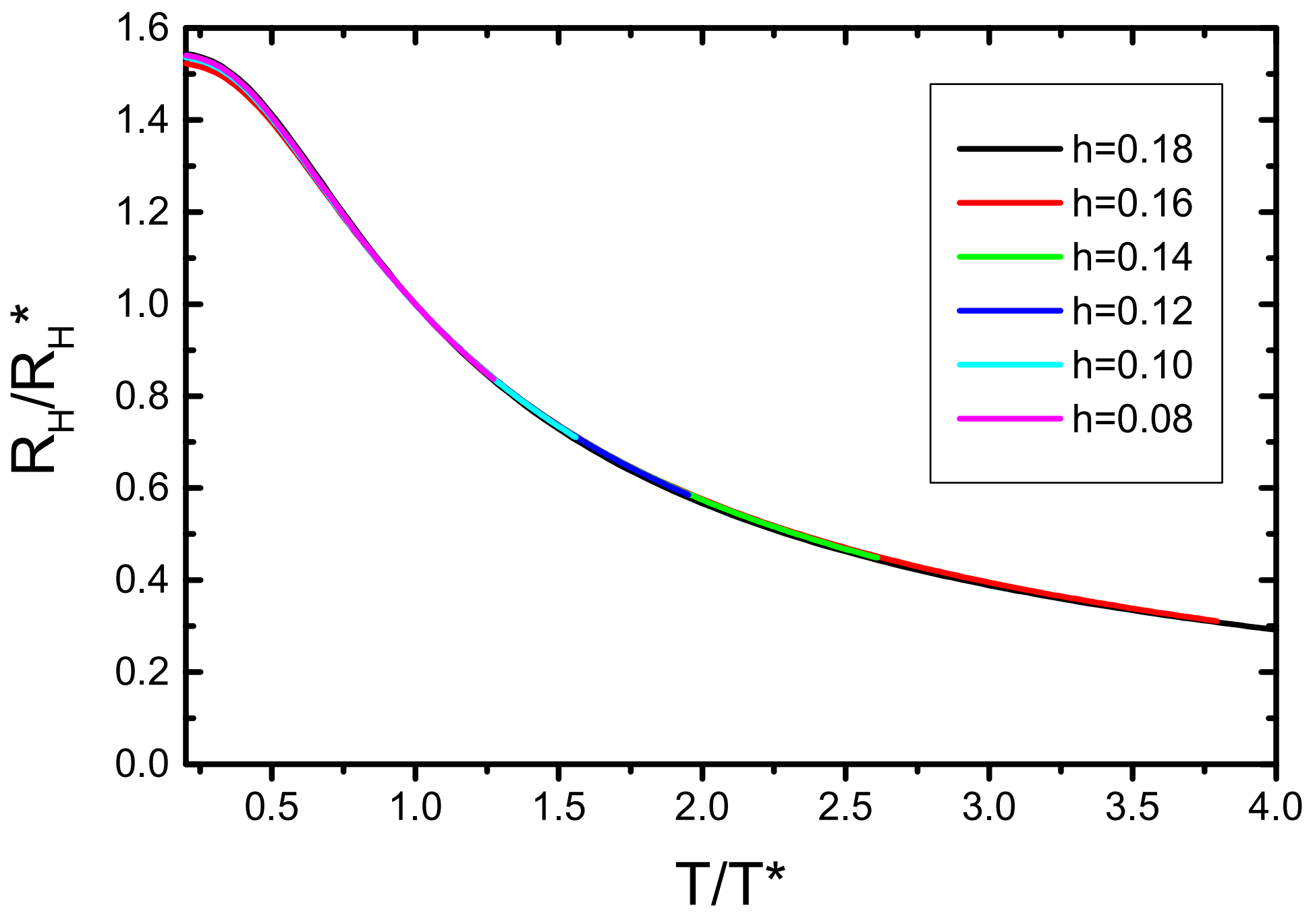

Figure 6: Universal curve for the ratio of $R_H/R_H*$ (here we cannot have true value for $R_H\infty$).

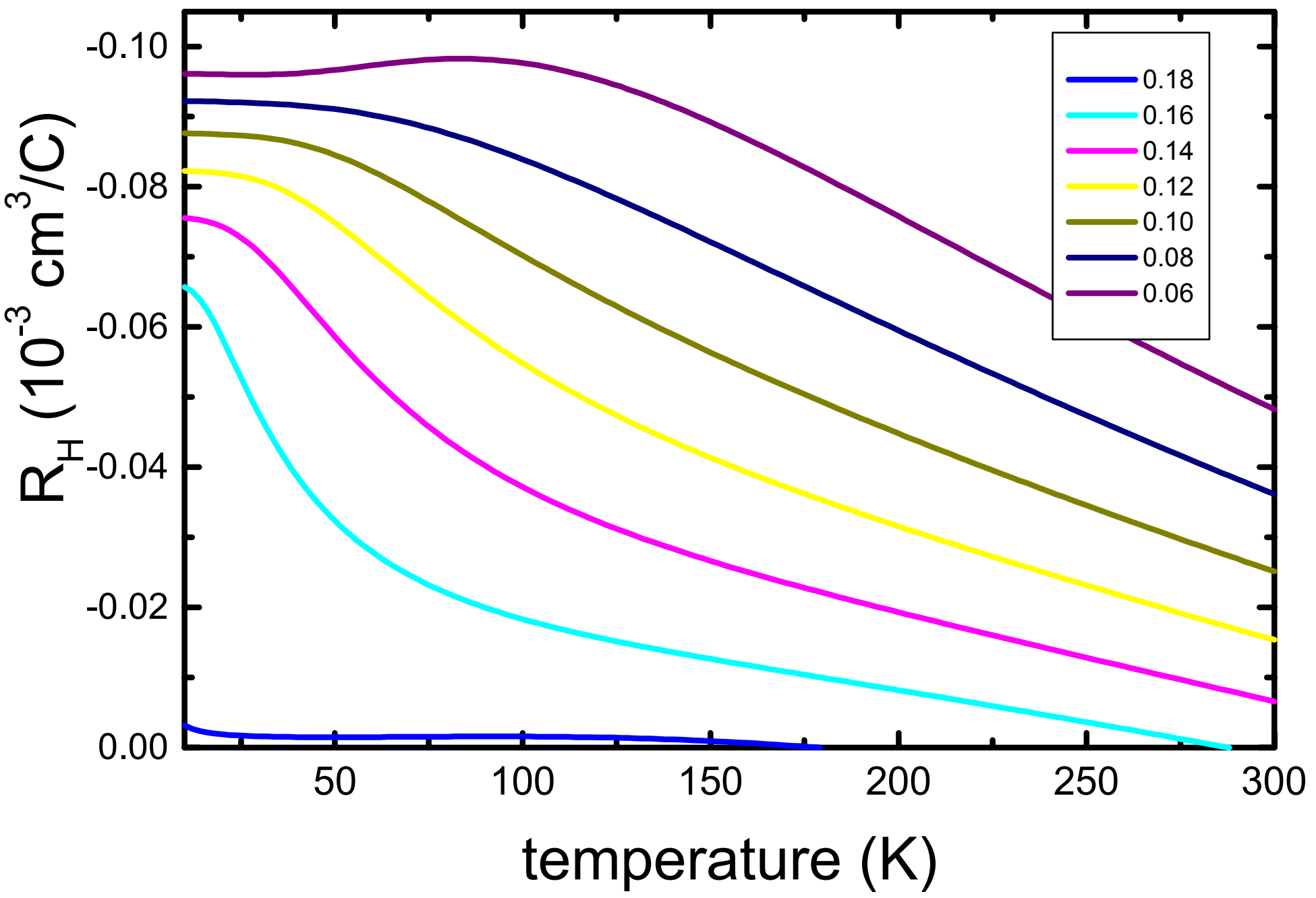

Figure 7: Calculated Hall coefficient $R_H$ for electron doping using formula [5].
With t = 0.18 eV, t' = 0.06057 eV, 2t'/t = 0.673